\documentclass[notitlepage,aps,pra,twocolumn,groupaddress,10pt]{revtex4-1}

\usepackage[english]{babel}

\usepackage{graphicx}
\usepackage[colorlinks=true, allcolors=blue]{hyperref}

\usepackage{stmaryrd}
\usepackage{amssymb,amsmath,amsthm,amsfonts,amsbsy}
\usepackage{bm,bibunits,color,chngcntr,epsfig,epstopdf,graphicx,dsfont}
\usepackage{hyperref,lipsum,,makecell,mathrsfs,rotating,setspace}
\usepackage[english]{babel}
\usepackage[normalem]{ulem}

\newcommand{\be}{\begin{equation}}
\newcommand{\ee}{\end{equation}}
\newcommand{\bea}{\begin{eqnarray}}
\newcommand{\eea}{\end{eqnarray}}
\newcommand{\ket}{\rangle}
\newcommand{\bra}{\langle}

\newcommand{\I}{\mathds{1}}
\newcommand{\ra}{\rightarrow}

\newcommand{\ba}{\begin{align}}
\newcommand{\ea}{\end{align}}

\def\C#1{\mathcal #1}

\usepackage{tikz}
\usetikzlibrary{shapes,arrows,positioning,fit,calc}

\begin{document}

\title{All about quantum error correction: distillation, mitigation, self-correction and beyond}
\author{Dong-Sheng Wang}
\affiliation{Institute of Theoretical Physics, Chinese Academy of Sciences, Beijing 100190, China \\
School of Physical Sciences, University of Chinese Academy of Sciences, Beijing 100049, China}

\newtheorem{theorem}{Theorem}
\newtheorem{prop}[theorem]{Proposition}
\newtheorem{corollary}[theorem]{Corollary}
\newtheorem{open problem}[theorem]{Open Problem}
\newtheorem{conjecture}[theorem]{Conjecture}
\newtheorem{definition}{Definition}
\newtheorem{remark}{Remark}
\newtheorem{example}{Example}
\newtheorem{task}{Task}
\newtheorem{protocol}{Protocol}

\begin{abstract}
In this work, it is shown that many quantum error-manipulating techniques, 
such as distillation, error mitigation, and dynamical decoupling, 
are special cases of the most general framework for quantum error correction.
This unifying perspective is achieved by extending quantum error correction 
to include state-adaptive and channel-adaptive settings, as well as multi-stage coding scenarios. 
Based on this insight, a model of self-correcting quantum memory is also proposed. 
This work clarifies the relationship among these techniques and illustrates, 
through explicit constructions, how the unified perspective 
can guide the design of reliable quantum information systems.
\end{abstract}
\date{\today}

\maketitle

\begin{spacing}{1.2}

\section{Introduction}

As quantum information is inherently fragile, 
quantum error correction (QEC) is indispensable 
for any large-scale quantum information processor~\cite{KL97}. 
The standard QEC framework exhibits several key features: 
it is composable, meaning that error-correction cycles 
can be freely interleaved with logical gates; 
it applies universally across different modalities of quantum information processing, 
including computing, communication, and metrology; 
and it is, in principle, efficient, or at least effective in practice, in order to be useful.
Most importantly, QEC is the central part of quantum Shannon theory
for information protection~\cite{Wat18,Wil17,Hay17}.

Despite this generality, a variety of other error-manipulating techniques,
such as state distillation, error mitigation, and dynamical decoupling,
have been developed largely in parallel with QEC~\cite{LB13}. 
These methods are often motivated by near-term constraints
and are typically not framed as instances of QEC, 
partly because their focus is not on constructing a good code or an optimal decoder. 
However, this fragmentation of the field tends to obscure the fundamental principle 
underlying all such protocols. 
In this work, we adopt a unifying perspective grounded in quantum Shannon theory
and show that QEC, in its most general formulation, 
serves as the underlying language for all these tasks. 
Specifically, we demonstrate that distillation, 
error mitigation, and dynamical decoupling can each be understood 
as special cases of state‑adaptive or channel‑adaptive QEC.

Although standard QEC is asymptotically efficient, 
the overhead required for fault-tolerant implementation 
can be prohibitive for near-term devices~\cite{Pre18}. 
Consequently, several relaxed forms of QEC have been proposed 
that scale less favorably but can still be effective in practice~\cite{WZO+20,PSB+21,AMO+24,KGC+24}. 
Understanding these practical variations within the unified QEC framework 
is one of the central goals of this paper, 
as it provides a systematic way to assess their resource trade-offs and design principles.

This paper is organized as follows. 
In Sec.~\ref{sec:int}, we review the framework of quantum error correction,
emphasizing its state-adaptive and channel-adaptive extensions 
as well as the Petz recovery map~\cite{OP93} as a universal decoder. 
Section~\ref{sec:spe} then recasts several well-known quantum techniques
including state distillation, stabilizer codes, passive Hamiltonian protection,
and error mitigation as special cases of QEC, 
thereby demonstrating the breadth of the unifying perspective. 
Section~\ref{sec:sa} develops a concrete model of self-correcting quantum memory 
based on state-adaptive repetition codes and AKLT-type systems~\cite{AKLT87}.
Finally, Sec.~\ref{sec:conc} concludes with a summary and an outlook on future directions.

\section{Quantum error correction}\label{sec:int}

In this section, we present the theoretical framework of quantum error correction (QEC).
A seminal result is that for a logical space that carries the 
quantum information, called a QEC code 
which is specified by a projector $P$, 
a set of error operators $\{E_i\}$ acting on it is correctable 
iff the following condition is satisfied 
\be PE_i^\dagger E_j P= c_{ij}P. \label{eq:qec}\ee 
Meanwhile, the set $\{E_i^\dagger E_j\}$ is said to be detectable~\cite{KL97}. 
If one chooses an orthonormal basis $\{|\psi_a\ket\}$ for the code space,
the condition above reads 
\be \bra \psi_a|E_i^\dagger E_j|\psi_b\ket=c_{ij}\delta_{ab}.\ee
It means the operators $E_i^\dagger E_j$ do not carry any information of a state $|\psi_a\ket$.
An operator that is not detectable will make a logical action on the code,
and the logical action may not be unitary. 
By definition, a unitary operator $U$ is a logical gate iff $[U,P]=0$.

How to detect a set of errors? It is from the QEC conditions above.
A detection scheme can be constructed as the measurement with $P$ and its complement $Q=\I-P$.
If the code starts from a logical state $\rho$,
disturbed by some error $E$ or not, 
this can be detected: if an error occurs, 
there will be some probabilities to realize $Q$,
if not, $P$ occurs with unit probability.
This basically reveals the direct-sum division of the whole Hilbert space $\C H=\C C \oplus \C C'$
into the code space $\C C$ and the `syndrome' space $\C C'$.
Any leakage into $\C C'$ makes an error detectable.

For error correction, the operation is more involved.
A basic scheme starts from the diagonalization of the condition~(\ref{eq:qec}) as 
\be PF_k^\dagger F_l P =\delta_{kl}d_k P,\ee 
for $d_k$ as the eigenvalues of the matrix $[c_{ij}]$,
which is interpreted as the state of the environment that induces the errors.
Then a recovery channel $\C D$ is constructed including the operators 
$R_k=\frac{1}{\sqrt{d_k}}PF_k^\dagger$:
given the error set $\{F_l\}$, which represents the same map as $\{E_i\}$,
the recovery can restore the original state with probability $p=\sum_k d_k$.
It is 1 if the set $\{E_i\}$ is a channel, namely,
completely positive and trace preserving~\cite{NC00}.
Closer analysis reveals that the effective action of a correctable error 
is unitary $\frac{1}{\sqrt{d_k}}PF_k^\dagger=PU_k^\dagger$, and they are orthogonal
\be PU_k^\dagger U_l P= \delta_{kl}P. \label{eq:u} \ee 
However, the unitary operators $U_k$ may not be local anymore even if $E_i$ are. 

\subsection{State-adaptive QEC}

If the protected state is known, 
a state-adaptive (SA) scenario is possible~\cite{W26}. 
Here we briefly recall its implication for error detection and correction.
For error detection, if an initial state is known, say a pure state $|\psi\ket$,
then a more fine-grained measurement can be used 
which is a projective measurement including $|\psi\ket$:
any non-zero probability on states other than $|\psi\ket$ indicates an error
or even a logical gate,
the latter would constrain the probability within the code space $\C C$. 
For error correction,
one also replaces the projector $P$ by the more fine-grained projectors 
$P_a$ for each basis state $|\psi_a\ket$,
and the recovery operators are $R_{ak}=P_a U_k^\dagger$
instead of $R_k$.
This would be modified if approximation is inevitable. 

A primary feature of the state-adaptive setting is that 
the recovery operation becomes inherently nonlinear, 
as it depends explicitly on the initial state. 
This nonlinearity is not a drawback; 
it arises naturally in many quantum information processing scenarios. 
For instance, in quantum computing one often starts from a known state 
and applies a predetermined sequence of gates, 
a task best understood as state generation. 
Although the knowledge of the state may not always be easy to exploit, 
admitting this dependence allows us to encompass state distillation 
and other protocols within the framework of state-adaptive QEC.

\subsection{Channel-adaptive QEC}

Besides the linear condition of the input state,
there is also a linear condition of the error set,
usually known as the linear Kraus-span (LKS) property~\cite{NC00}.
It is clear to see from the condition~(\ref{eq:qec})
any other set of operators with each as a linear combination of $E_i$ also 
satisfies the condition.
For instance, one often considers local on-site Pauli errors on physical qubits,
and if they are correctable, then any other form of on-site errors 
is also correctable.

The seminal Shannon theory for error correction is channel-adaptive, however~\cite{Wat18,Wil17,Hay17}.
This would violate the LKS property.
On the other hand, the benefit is also apparent:
it can enhance the error-correction performance.
This also naturally leads to approximate QEC which fits into practical settings.
Namely, some errors cannot be corrected, and the exact condition~(\ref{eq:qec})
is normally satisfied only approximately.

Using superchannel~\cite{CDP08a,WLWL24},
a QEC task can be defined as a superchannel $\hat{\C S}$ acting on $n$ parallel action
of a channel $\Phi$ so that 
\be F(\omega^{\otimes k},(\hat{\C S} (\Phi^{\otimes n})\otimes \I^{\otimes k})(\omega^{\otimes k}))
\geq 1-\epsilon, \label{eq:codeerror}\ee 
with $\epsilon\in [0,1]$ and
the state fidelity function between ebits $\omega^{\otimes k}$ (Bell states)
and the Choi state of $\hat{\C S} (\Phi^{\otimes n})$~\cite{Cho75},
for 
$F(\rho,\sigma):=\|\sqrt{\rho}\sqrt{\sigma}\|_1^2$,
with $\|\cdot\|_1$ denoting the trace norm.
In order to make a code useful, one has to require $\epsilon$ to be as small as possible,
and such approximate code is called `quasi-exact' code~\cite{WZO+20}.
The supremum of all achievable coding rate $\alpha:={k}/{n}$ is known as 
the quantum capacity of the channel.
Many types are possible by considering different types of coding $\hat{\C S}$
and even the fidelity function~\cite{Wat18,Wil17,Hay17}.

A central result is that in this setting a near-optimal universal recovery scheme is the 
Petz map~\cite{OP93,Wil17,BK02,NM10,BDL16}.
It also includes the exact setting above as a special case.
Given a channel $\Phi$ with its Kraus operators $K_i$, and also a state $\sigma$,
the standard Petz map $\C P_{\sigma,\Phi}$ is defined by a set of Kraus operators
\be R_i=\sigma^{1/2} K_i^\dagger \Sigma, \ee
for $\Sigma:=\Phi(\sigma)^{-1/2}$,
so that $\C P_{\sigma,\Phi}$ can be the optimal recovery on $ \Phi (\rho)$. 
The map itself is not trace preserving but it indeed is on the support of $ \Phi (\rho)$.

Special forms of it are often used in error correction, 
including the pretty-good measurement and transpose channel which uses a code projector $P$
instead of $\sigma$ for its Kraus operators~\cite{BK02,NM10}.
Actually, the Petz map $\C P_{\rho,\Phi}$ can fully recover a known
input state $\rho$ with $\C P_{\rho,\Phi}\Phi(\rho)=\rho$,
and this has been used to define the SA quantum channel capacity~\cite{W26}.

\subsection{Multi-stage QEC}

The noise channel may not occur in parallel,
and this scenario can be described by the quantum comb framework,
which is an extension of superchannel~\cite{CDP08a}.
By abuse of notation, denote such usage of channels amortized in spacetime 
as $\Phi^{n}$ and the coding comb still as $\hat{\C S}$,
and then the fidelity condition becomes
\be F(\omega^{\otimes k},(\hat{\C S} (\Phi^{ n})\otimes \I^{\otimes k})(\omega^{\otimes k}))
\geq 1-\epsilon. \label{eq:codeerror2}\ee 
It is largely unknown how this can be used for better QEC, however.

It turns out the early strategy of dynamical decoupling (DD)~\cite{VKL99} 
can be viewed as a special case of this. 
Namely, the corrected noises are just Pauli errors,
and the `comb' is formed by unitary operators on the system itself
induced by external control field. 
As no spatial overhead is used, the efficiency of DD is also seriously constrained. 
Improved DD with error-correction codes has also be studied~\cite{LB13}.

Quantum comb has found applications in other tasks such as channel discrimination and 
estimation~\cite{CAP08}.
These tasks also relate to QEC. 
For instance, the swap-test~\cite{BCW+01} can discriminate between states, 
and it is recently found that it can also be used for QEC~\cite{CFL+25,Koc21},
as we will see in Sec.~\ref{sec:sa}.

\section{Special cases of QEC}\label{sec:spe}

In this section, some special cases of QEC are studied.
In order to properly interpret them, 
we first use quantum resource theory (QRT) to sharpen the framework of QEC, 
hence also quantum Shannon theory. 
Briefly, a QRT divides a set of objects into a free set $\C F$
and a resource set $\C R$, and $\C F$ is defined by free operations $\C O$ that preserve it~\cite{CG19}.
A QRT framework for QEC has been developed recently~\cite{WLWL24} by 
considering the set of coding superchannels,
with a channel capacity as the maximal resource measure of a code.
Resourceful codings are those that can increase the coding fidelity~(\ref{eq:codeerror}).

Codings act on channels,
and a level lower is the QRT of channels, which further act on states.
An example is to consider local operations with clsssical communication (LOCC) as the free set, 
and any entanglement-generating channel becomes resourceful.
For resources of states,
a seminal example is entanglement,
with separable states as the free set and LOCC
as free operations,
and the resource is entanglement~\cite{HHH+09}.

A primary task in a QRT is the resource preparation task.
For QEC, this becomes logical state preparation. 
While in literature
QEC is often viewed as the channel distillation of $\I^{\otimes k}$ from $\Phi^{\otimes n}$
by free operations~\cite{DHW08,ADHW09}.
It is easy to see the two point of views are actually equivalent.
As coding acts on channels, 
a resourceful code can be viewed as the yield of a free code acting on a resourceful channel.
Similarly, a resourceful state can be viewed as the yield of 
a free channel acting on a resource state. 
Therefore, a resource \emph{generation} task can be viewed 
as a resource \emph{conversion} task on a lower level, and vice versa. 
The resource distillation is a seminal example of resource conversion.

Below we analyze a few examples of QEC. 
One is the usual state distillation task, 
and the goal is to make clear that 
it aligns with the SA QEC;
and the second is the QEC with stabilizer codes,
and the goal is to show that the way we use stabilizer codes
is only part of the whole paradigm of QEC.
It also includes passive QEC with Hamiltonian protection,
which is relatively less explored and relates to self-correcting quantum memory.
Finally, we show that error mitigation is a special case of QEC.

\subsection{State distillation}

A seminal example is entanglement distillation which is to distill ebits from copies of a state
\be \rho^{\otimes n} \ra \omega^{\otimes k}, \ee 
only allowing free LOCC.
The rate $r=k/n$ is known as the distillation rate~\cite{DW04,DW05}.
Another example is magic-state distillation~\cite{BK05} 
which is to distill `mabit' $|t\ket=TH|0\ket$ from copies of a state
with Clifford operations~\cite{GC99}.

It is well known that QEC codes can be used for state distillation,
and it is easy to see these are SA QEC.
There is no channel in the usual distillation task, 
yet a state $\rho$ can be viewed as the output from a channel $\Phi$
acting on a fiducial state.
For a known state, multiple copies of it can be prepared
as a semi-classical repetition code,
which is impossible if the state is unknown,
consistent with the no-cloning theorem. 
The usage of copies of states is also widely used in other tasks such as error mitigation
and tomography~\cite{NC00}.

\subsection{Stabilizer codes}\label{subsec:sta}

An important class of QEC codes is the class of stabilizer codes~\cite{Got98,NC00},
and due to its success, 
it is somehow often viewed as the synonym of QEC codes.
A stabilizer code is defined by a set of commuting stabilizers,
which can be used for the encoding and decoding of the code.
It can also be slightly extended to subsystem codes with non-commuting gauge operators,
but still commuting stabilizers~\cite{Pou05}.
Often qubit stabilizer codes are employed
with each stabilizer as a tensor-product of Pauli operators
and only with eigenvalues $\pm 1$.

Stabilizer codes are often designed as exact codes for Pauli errors.
Given stabilizers $S_i$,
the QEC with a stabilizer code often uses projective measurements 
\be P_i=(\I\pm S_i)/2.\ee
A Pauli error can only commute or anti-commute with a stabilizer,
so that a correction operator $P_iU_k$ from~(\ref{eq:u}) becomes $U_k P_j$,
wherein a stabilizer projector $P_j$ can be performed first,
and then a Pauli correction $U_k$ is applied.
For error detection, one can simply measure some stabilizers 
without applying Pauli correction. 
Stabilizers are often local, making then easy to be measured.

Stabilizer codes used as above are neither state adaptive, nor channel adaptive.
However, they still can be used in these ways.
If a codeword is known to be a stabilizer state, 
stabilizers of it can be used to correct more errors than merely 
using the stabilizers of the code~\cite{W26}.
One can also consider the correction of other errors with stabilizer codes,
for instance erasure errors and amplitude damping channels~\cite{NC00}.
It is often easier to correct erasure errors than Pauli errors,
and correcting two erasures is equivalent to correcting one Pauli error~\cite{GBP97,BDS97}.
In the condition~(\ref{eq:qec}), the matrix $[c_{ij}]$ needs to be diagonalized,
and the effective operators $F_k$ may not be local anymore.
This is the case for amplitude damping.
A way to avoid the nonlocal correction is to first use the encoding 
$|0\ket=|01\ket$, $|1\ket=|10\ket$, which converts amplitude damping into erasure~\cite{W26},
and then use a stabilizer code to correct erasure. 
Other types of codes are also known to correct them such as the permutation-invariant codes~\cite{OC20}.


\subsection{Passive protection}

A code can be viewed as the ground subspace of a quantum Hamiltonian.
The seminal example is Kitaev's toric code~\cite{Kit03},
which can be viewed as a code with active QEC,
or a ground-space code. 
The passive protection is due to the finite gap to excited states,
and cooling the system effectively realizes QEC.

As a ground subspace, the encoding can use the universal feature of a phase of matter.
Namely, for a parameterized model $H(\lambda)$,
a gapped phase of matter could exist for some values of $\lambda$.
The topological codes (including the toric code)
and symmetry-protected topological (SPT) codes are such examples~\cite{ZCZ+15}. 
The code space $\C C$ contains a `gauge' part that specifies a particular state
and a logical part that is universal for a phase. 
The gauge part does not matter for the logical actions.

The QEC condition~(\ref{eq:qec}) now reads as a correlation function.
The correction can be described as the cooling $e^{-\beta H}$ 
together with thermal jump operators for very low temperature $1/\beta$~\cite{Bre03}.
The temperature controls the average number of excitations, i.e. errors,
and by maintaining a low temperature,
there is no need to annihilate them all.
However, logical gates could create more excitations as they do not commute in general.
Therefore, cooling is necessary between logical gates to maintain 
a small number of excitations.

\subsection{Error mitigation}

Quantum error mitigation (QEM) has been developed recently as an effective set of techniques
to reduce errors, and can also be combined with QEC codes~\cite{PSB+21,SEF+22}. 
In this section, we will show that QEM, in the most general sense of it,
is a special type of QEC,
just as other examples shown above. 


Instead of operators, QEM focuses on classical values, namely, 
expectation value of observable. 
This violates the composability of usual QEC since 
a state after measurement cannot be used further.
Yet, mathematically, a state $\rho$ can be reduced to a set of 
classical values as 
\be \rho=\sum_i  A_i^\dagger \; \text{tr}(A_i\rho), \ee 
if $\{A_i\}$ forms an operator basis. 
Each classical value $\text{tr}(A_i\rho)$ is further reduced
to the estimation of measurement probabilities.
Computing expectation value rather than preparing a state 
would not reduce the computational complexity~\cite{W15} 
if the observable is arbitrary as this can be used to reconstruct the state.

A general framework of QEM and its cost have been studied lately~\cite{TEM+22}.
Here it will be presented as a type of SA QEC. 
As the initial state $\rho$ is known, multiple copies can be prepared.
A noise channel $\Phi$ to be corrected acting on $\rho$ is also known,
and can be manipulated to other forms.
The goal is to find the effective inverse of $\Phi$ restricted onto $\rho$.
While the SA QEC achieves this goal with Petz map as 
the general decoding scheme, 
here QEM is a restricted inefficient way since it does not use 
entangling encoding.
In some QEM the observable $A$ is also not arbitrary,
but here we assume it is. 

First of all, without noise the sampling cost $M$ scales exponentially 
with the accuracy $\nu$ for $M\sim \frac{1}{\delta^2}$, and $\delta \sim 2^{-\nu}$
according to the standard Hoeffding's inequality~\cite{Hay17}.
This can only be quadratically better if the quantum amplitude estimation 
algorithm is used~\cite{BHM02}.
Often, this cost is not a major concern in practice.

With noise, a $(Q,K)$ mitigation scheme is able to perform joint measurement on $Q$ copies of samples,
forming a cluster,
and use $K$ clusters each with a different modulated noisy channel other than $\Phi$ itself, 
and post-process the measurement outcomes.
The mitigation cost is assessed by a maximum bias parameter $b$ and a maximum spread $\Delta$
which is the difference between the maximum and minimum
possible estimated expectation values for all possible $A$ and $\rho$.
It is found that the sampling cost is $\C O(\Delta^2 /\delta^2)$,
and $\Delta$ is lower bounded by a value based on distance between states~\cite{TEM+22}.
The reason is that quantum channel would reduce the difference between states~\cite{NC00},
therefore, mitigation needs to slow information blurring due to noise.
For specific QEM, $\Delta$ would scale exponentially with the circuit size,
e.g., the number of qubits or the circuit depth,
rendering it inefficient.

Despite being inefficient, 
it can be effective in practice for small circuits.
This has been obvious from the development of DD,
which has been a seminal module in quantum control of errors~\cite{LB13}.
In order to reduce the cost of standard QEC codes and increase the accuracy of DD,
concatenation between them can be employed.
There are two primary schemes:
\begin{itemize}
    \item Inner DD with outer code: each physical qubit can be manipulated by DD 
    to reduce a certain type of noises, e.g., phase-flip errors, 
    and the outer layer code is used to correct bit-flip errors.
    \item Inner code with outer DD: use a small code that only corrects some noises,
and use DD sequence of error-correction cycles to correct the rest.
\end{itemize}


As the case of DD, QEM and SA QEC in general, can be concatenated with QEC codes.
Here we layout the two primary types of concatenation.

It is straightforward to use a QEC code as the inner code and SA QEC code as the outer code. 
For this setting, small QEC codes are sufficient as otherwise there is no need to use the outer code.
The outer code is responsible for the leftover errors from the inner codes. 
Also non-composable QEM schemes can be used if all the logical gates has been done in the inner code.

If a SA QEC code is taken as the inner code, it has to be composable, meaning that 
a protected state has to be available for further quantum processing.
There shall be an advantage for using it instead of a usual QEC code,
e.g., regarding encoding cost or noise type. 
An example is to use distillation or mitigation for the magic state~\cite{PSB+21},
injected to a stabilizer code.
Compared with the coding switching~\cite{BKS21}, 
a full even assessment of their advantages would depend on many factors. 

\section{SA QEC codes}\label{sec:sa}

In this section, we study SA QEC codes and the constructions of them
are far from clear at present.
Despite this, from very primary SA QEC codes 
we propose a class of SA self-correcting quantum memory 
based on an ensemble code.

\subsection{General features}

We first analyze general features of SA QEC codes.
Due to the COPY operation, such codes can be nonlinear, 
and we would denote a SA QEC code as \be \{n,k,d\} \ee
for encoding $k$ qubits into $n$ qubits, with code distance $d=2t+1$,
for $t$ as the maximal number of correctable noise of a certain type.
Entangling operations at the encoding as well as the decoding are allowed.
If local parity checks are available, the fault-tolerance of the decoding 
can be achieved by measuring them repeatedly.



There are interesting interplay with usual QEC codes.
A SA code, as a manifold, can be embedded into a linear QEC code,
hence constructed from it. 
Meanwhile, a usual QEC code can also be used in a SA fashion,
as shown in Sec.~\ref{subsec:sta}.
Moreover, SA QEC codes do not obey the usual condition~(\ref{eq:qec}), 
so they could violate quantum coding bounds but still obey the classical bounds. 
For instance, they obey the Singleton bound 
\be d-1\leq n-k, \ee
instead of the quantum version $2(d-1)\leq n-k$.
Below it is shown that the smallest such code has size $n=3$ instead of 5.

It is expected that SA type LDPC codes can be constructed
by specifying a set of parity checks.
Recall that a LDPC code is defined with a finite set of commuting parity checks,
with each check only acting on a finite number of bits,
and each bit only acted upon by a finite number of checks~\cite{RL09}.
The stabilizer LDPC codes employ Pauli stabilizers as commuting checks,
and errors collapse to Pauli errors that commute or anti-commute 
with a stabilizer. 
For the SA case, the checks may not commute.
For instance, SWAP operations, as elements of the permutation group $S_n$, 
can serve as checks yet they do not commute.




\subsection{SA quantum repetition codes}

The simplest type of classical code may be the repetition codes. 
A basic example is to encode a bit value 0 as 000, and 1 as 111.
It only has a single logical gate, the Pauli bit-flip gate $X_L=XXX$.
It easily extends to a code $[n,1,n]$ with a large code distance $d=n$.
Yet there are two basic types of decoding:
one is based on parity check, which is to find the values of 
the parities $ZZ$ on each nearest neighbor pair,
and use majority vote to find the bit-flip errors and correct them;
the other is based on a global `order check', 
which is to find the value of $\sum_i Z_i$, 
and then use a global field to polarize this value.
Namely, it does not identify each error, 
instead, it reduces the number of errors on average. 

There are two primary settings where classical repetition codes apply.
One is in communication, and one is in memory (storage drive). 
The magnetic hard drive, which is described by the Ising model 
and in which a bit is represented by a spin, 
uses the classical repetition code.
It inherently poses the parity checks as its Hamiltonian
$H=-\sum_{i}Z_i Z_{i+1}$,
so that only the global order check is needed. 
Actually, this system is \emph{self-correcting}, i.e.,
no active error correction is needed as long as 
the system remains in the ferromagnetic phase below the phase transition
temperature~\cite{BLP+16}.
The encoding rate can be larger by using more magnetic domains
instead of the ground states. 
The code is also expanded to a subsystem code,
with 0 or 1 as the total magnetic order of a domain,
rather than a single configuration within it. 
Any thermal fluctuation below phase transition would not matter.

\subsubsection{Self-correcting SA quantum memory}

We now construct the SA quantum repetition codes, 
named as `ensemble codes' for short.
The key point is to realize that the analog of a bit is a qubit
at a known state. 
A simple ensemble code just maps a state $|\psi\ket$ into 
a few copies of it $|\psi\ket^{\otimes n}$.
Here, the COPY is a valid operation since the state is assumed to be known,
consistent with the no-cloning theorem.
We assume the state is pure but can also be a mixed state,
and a codeword is denoted as $\psi^{\otimes n}$.
The code has to be SU(2) covariant to store any state $\psi$.
In order to enable a nontrivial Hamiltonian protection,
we have to consider systems with SU(2) symmetry,
which are nothing but the AKLT systems~\cite{AKLT87}. 

In this system, a state $\psi$ is not stored as a qubit,
instead, it is stored as a Choi state~\cite{Cho75}. 
Namely, $\psi$ is firstly identified with a qubit unitary gate $U$,
which is then stored as its Choi state 
\be |U\ket=(U\otimes \I)|\omega\ket,\ee
and $|\omega\ket=(|00\ket+|11\ket)/\sqrt{2}$ is the ebit.
Such a storage can also be viewed as the storage of the parameters in a gate $U$
instead of a qubit. 
An AKLT model can be defined on any dimension or lattice,
while the common element is the `valence-bond' picture.
A valence bond is the singlet state $(|01\ket-|10\ket)/\sqrt{2}$,
which is just the ebit up to a Pauli correction. 
Given a collection of valence bonds $\omega^{\otimes n}$,
a ground state is constructed by applying local projectors 
to spin sectors.
A parent Hamiltonian that takes this state as the unique ground state
can be found, and importantly, this system has SU(2)  
(or SO(3) equivalently) SPT order~\cite{ZCZ+15},
and it has been used to define a family of QEC codes~\cite{WAR18}.

\begin{figure}[t!]
    \centering
    \includegraphics[width=0.45\textwidth]{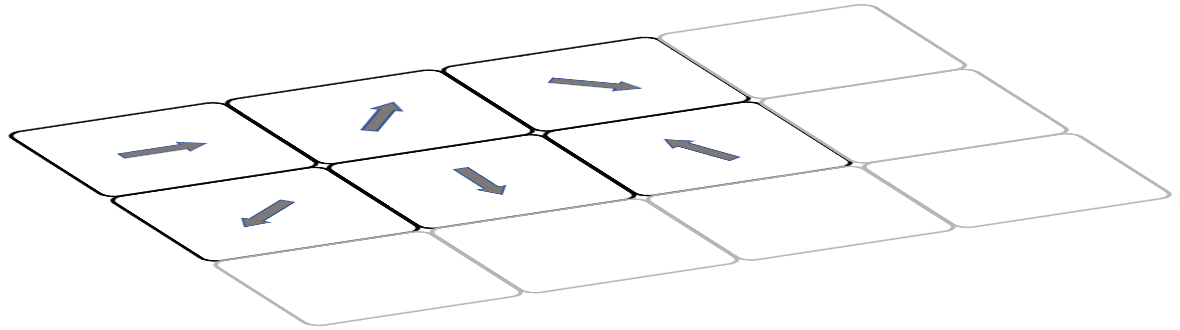}
    \caption{An illustration of the ensemble code using 3D AKLT system.
    The picture is the top of such a system with each vector encoding a SU(2) gate,
    as a generalization of the Ising model as classical repetition code
    which only encodes two logical vector directions.
    }
    \label{fig:aklt_memory}
\end{figure}

Instead of the ground state, we now use domain walls to encode qubits.
Note, rather than domains it uses domain walls,
which is a manifest of the holographic bulk-edge duality,
and the dilute antiferromagnetic order of the system.
The `domain wall' here is at the valence bond.
We apply a global gate of the form 
\be U_1^{\otimes l_1} U_2^{\otimes l_2} \cdots U_k^{\otimes l_k} \ee 
with each $U_i\in$ SU(2) and $l_i$ as the length of a `symmetry twist'.
Due to the global symmetry, there will be 
a nontrivial gate on each domain wall, creating excitations known as spinon or soliton.
This can be seen based on the matrix-product state representation of the state~\cite{W20_qubit}.
In order to increase the size of the domain wall,
one can consider 2D or 3D systems,
and a domain wall will be the shape of a circle or surface, respectively. 
We illustrate the code with the top of a 3D system, shown in Fig.~\ref{fig:aklt_memory}.
Each arrow for a gate $U_i$ lives only on the surface of a 3D domain.
If using the Bloch vector of qubit,
an intuitive picture of the code is to view it as a collection of vectors each 
pointing to an arbitrary direction. 
Note the vectors live in the `virtual' valence-bond picture,
but they effectively behave as real spins.
To ensure fault-tolerance, finite-digit precision can be enforced 
on the continuous parameters to be stored. 
In comparison, the classical case only allows two directions: up and down.

Such a system, although simple enough, could be self-correcting.
Similar with the Ising model, 
there is no self-correction in 1D since a domain wall is of zero dimension 
and a domain can grow without energy cost.
But in 2D and 3D, a domain wall has a size.
For Ising model, the growth of its size will cause energy so it is confined,
leading to the finite-temperature 2nd order phase transition from a random paramagnetic phase.
For the chosen AKLT models, as the domain-wall gates do not commute 
with the Hamiltonian terms at the domain wall, 
the change of domain wall size also causes energy. 
Besides, each spin vector in a domain cannot easily deviate from its logical direction
since that will violate the symmetry.
At high-enough temperature, thermal fluctuations would randomize each logical vector,
erasing logical information.

A self-correcting QEC code is the hypothetical 4D toric code~\cite{DKL+02},
for which the two types of excitations are both size-confined.
By poking holes it can store more qubits, but storing multi-qubit entangled states 
are challenging as that requires applying entangling logical gates,
similar with our AKLT ensemble code. 
There are no-go theorems~\cite{BLP+16} regarding the self-correction of quantum
LDPC stabilizer codes, and these codes in principle are expected to store entangled states. 
However, as a memory this may be unnecessary and storing qubit states could be sufficient
in practice. 

Realizing our ensemble code is challenging, though. 
First of all, AKLT models, although mathematically concise and with almost half a century history,
have seldom been realized in quantum materials.
They are strongly correlated systems and have strict requirements on symmetry.
Secondly, whether a AKLT model is gapped or not is a hard problem
as it depends not only on the dimension, but also on the geometry and spin values.
Finally, the read/write operation on the code could be hard.
The magnetic direction of each domain has to be readable and writable.
This process also has to be accurate and fast. 

\subsubsection{Parity check}

In this section, we study ensemble code further, 
but instead of using global order check,
we study how to use local parity check. 
First of all, we have to figure out what is the `parity' in this setting.
For AKLT systems, the local parity is nothing but the local Hamiltonian interaction terms,
which are well defined yet not easy to measure.
Here, we study the setting when a state $\psi$ is 
directly encoded as $\psi^{\otimes n}$.

It is clear to see its parity check is nearest-neighbor swap gates.
A noise channel $\Phi$ may induce error on some of the local $\psi$s.
A parity check is just the well-known swap-test~\cite{BCW+01}:
it requires a qubit ancilla to perform the projectors $P_\pm= (\I\pm \text{SWAP})/2$,
the outcome $+$ means no error, and $-$ means the two states are not the same,
hence error occurs. 
Yet, the location of the error has to be found, 
therefore, one needs $n\geq 3$ to achieve this, 
and there are $n-1$ parity checks, and this is a code $\{n,1,n\}$.
This is very similar with the classical repetition code $[n,1,n]$
by treating $\psi$ as the analog of bit value 0.

How to correct the errors? In the classical setting, one applies bit-flips $X$
and sometimes there is even no need to do the correction.
By majority-vote and omitting these erroneous bits below the error threshold,
the correct logical value can be decoded. 
In the quantum setting, it is similar. 
To actively correct the error, the channel $\Phi$ has to be known,
which is much more complicated than Pauli operators. 
For this purpose, one can find the pseudo-inverse of it, $\C D$, such that 
$\C D \Phi (\psi)\approx\psi$~\cite{CLK+21}. 
The map $\C D$ can depend both on $\psi$ and $\Phi$.

Another setting is that each qubit is affected by the same noise.
This also applies to the classical case, say,
a configuration of a repetition code would be $\rho^{\otimes n}$, 
for $\rho$ as the random variable which is a diagonal state diag$(p,1-p)$,
for $p$ as the bit-flip rate.
The quantum case merely adds a transversal rotation on it. 
In this setting, it is recently found that~\cite{CFL+25} 
upon two copies of a mixed state $\rho\otimes \rho$,
the swap-test post-selected on the $P_+$ outcome can increase the purity of $\rho$
against the noiseless state $\psi$ for $\rho=p \psi + (1-p) \sigma$,
for $\sigma$ as the noisy part. 
A sequence of $P_+$ will drive the state to the pure fixed point $\psi \otimes \psi$,
as a type of Zeno effect. 
For any $n$, measuring a $n$-element permutation operator in $S_n$ can also suppress 
the noisy part~\cite{Koc21}. 
The similar scheme for the classical case is to repeat the parity checks $ZZ$.
The post-selection is not efficient but could be effective in practice. 

For $n=3$, one can measure two swap operators which has four outcomes.
The local qubit state is one of the following forms: 
$(5+f)\rho+2\rho^2$, $(1-f)\rho$, $(1+f)\rho-2\rho^2$, 
for $f=\text{tr}(\rho^2)$ as the purity of $\rho$, modular normalization.
The first one is the good case with an enhanced purity and can be reused,
the second one with no state change can also be reused,
while the last one requires a correction $\C D$, which depends on $\rho$. 
For instance, if $|\psi\ket=U|0\ket$ and the error is a rotated bit-flip $UXU^\dagger$,
then the correction is simply $XU^\dagger$.
In general, the correction $\C D$ may not be unitary but can be easily found when the dimension of $\rho$ 
is small. 

\section{Conclusion}\label{sec:conc}

In this work, we have theoretically explored a unifying picture of quantum error correction (QEC), 
showing that other techniques such as error mitigation and state distillation 
can be understood as special cases of QEC, 
particularly within the state-adaptive (SA) framework developed recently. 
As a concrete example, 
we proposed a SA quantum memory that is expected to be self-correcting, 
based on encoding logical information in the polarization directions of spinon excitations in AKLT-type systems.

It is worth noting that both AKLT systems and cluster-state systems 
are known to provide universal resources for quantum computing~\cite{BBD+09}. 
We have previously shown that cluster states are suitable for constructing quantum transistors 
that serve as quantum internal memory or processing unit~\cite{LXW+26}. 
Here, we demonstrate that AKLT systems are naturally suited to serve as quantum external memory. 
This external memory, which we term the ``ensemble code,'' 
can be viewed as semi-classical and analog in nature, 
as it stores continuous parameters rather than discrete bit values. 
Importantly, this scheme does not violate the no-cloning theorem or the Holevo bound: 
it encodes a known qubit state as multiple copies, 
and the global state can be read out from macroscopic features. 

Our work also highlights the potential advantages of SA QEC codes 
and their intricate interplay with conventional QEC codes. 
The overhead of current stabilizer codes is substantial, 
largely due to the frequent refreshing of stabilizer measurements. 
We expect that SA codes can reduce this overhead by exploiting knowledge of the initial states, 
the noise channels, or the overall algorithmic function, 
while maintaining practical efficiency and efficacy.

\section{Acknowledgement}

This work has been funded by
the National Natural Science Foundation of China under Grants
12447101 and 12105343.

\end{spacing}


\bibliography{ext}{}
\bibliographystyle{elsarticle-num}

\end{document}